\documentclass{PoS}

\usepackage{graphicx}
\usepackage{amssymb,amsmath,amsfonts}
\usepackage{color}
\usepackage[utf8]{inputenc}
\usepackage[caption=false]{subfig}

\title{Possible higher order phase transition in large-$N$ gauge theory at finite temperature}

\ShortTitle{Possible higher order phase transition in large-$N$ gauge theory}

\author{\speaker{Hiromichi Nishimura}\thanks{H.N.~is  supported by the Special Postdoctoral Researchers program of RIKEN.} \\
        RIKEN/BNL Research Center, Brookhaven National Laboratory, Upton, NY 11973   \\
        E-mail: \email{hnishimura@bnl.gov}}

\author{Robert D. Pisarski\\        
	Department of Physics,  Brookhaven National Laboratory, Upton, NY 11973\\        
	E-mail: \email{pisarski@bnl.gov}}        

\author{Vladimir V. Skokov\\        
	RIKEN/BNL Research Center, Brookhaven National Laboratory, Upton, NY 11973 \\        
	E-mail: \email{vskokov@bnl.gov}}

\abstract{
We analyze the phase structure of $SU(\infty)$ gauge theory at finite temperature using matrix models.  Our basic assumption is that the effective potential is dominated by double-trace terms for the Polyakov loops.  As a function of the temperature, a background field for the Polyakov loop, and a quartic coupling, it exhibits a universal structure: in the large portion of the parameter space, there is a continuous phase transition analogous to the third-order phase transition of Gross, Witten and Wadia, but the order of phase transition can be higher than third.  We show that different confining potentials give rise to drastically different behavior of the eigenvalue density and the free energy.  Therefore lattice simulations at large $N$ could probe the order of phase transition and test our results.
}

\FullConference{Critical Point and Onset of Deconfinement - CPOD2017\\
		7-11 August, 2017\\
		The Wang Center, Stony Brook University, Stony Brook, NY}

\begin{document}

\section{Introduction}

The phase structure of $SU(N)$ gauge theories is of fundamental importance. 
In finite-temperature pure gauge theory, the relevant order parameter is the Polyakov loop. It is therefore reasonable to study the phase transition as a function of an effective theory of thermal Wilson lines, as a type of matrix model. 

At infinite $N$ the problem can be simplified by writing the effective potential as a functional of the eigenvalue density \cite{Brezin:1977sv}, rather than as a function of $N-1$ eigenvalues. The phase structure at infinite $N$ becomes also interesting as various models show an exotic phase transition whose order is third. Gross, Witten and Wadia first showed that in lattice gauge theory in two dimensions, there is a third-order phase transition as a function of the coupling constant \cite{Gross:1980he,Wadia:1980cp}.  Lattice gauge theory at strong coupling with heavy quarks also exhibits a third-order phase transition as a function of temperature \cite{Green:1983sd,Damgaard:1986mx}. 

In this paper, we study a phase structure of $SU(\infty)$ gauge theory with an external field of the Polyakov loop as a type of matrix model.  
We show that depending on the type of confining potential, there is a continuous phase transition whose order is third or higher.  This phase transition is a generalization of the Gross-Witten-Wadia (GWW) phase transition. 

\subsection{Polyakov loop at large $N$}

The Polyakov loop, $\mathrm{tr} \,L(\mathbf{x}) $ where 
$L(\mathbf{x})=
\mathcal{P} \exp \left[ {i g \int^{1/T}_{0} d\tau A_0 (\tau,\mathbf{x}) }\right] $, is a Wilson loop along the temporal direction $\tau$ in the
fundamental representation. 
Under center symmetry, the Polyakov loop transforms as $\mathrm{tr} L
\rightarrow z \mathrm{tr} L$ where $z \in Z(N)$. After diagonalizing the
Polyakov line by a gauge transformation, we denote the eigenvalues as $\theta_i$ where $L = \mbox{diag} (e^{i\theta_1}, \dots,e^{i
\theta_N} )$ with $\theta_1 + \theta_2 +\dots + \theta_N=0$. 
Since there are $N-1$ degrees of freedom in the group manifold of $SU(N)$,
the Polyakov loop alone is not sufficient to describe the theory for $N>2$. One can either take  $N-1$ independent eigenvalues or use the Polyakov loops that wind $n$ times in the temporal direction,
\begin{equation}
\rho_n = \frac{1}{N} \mathrm{tr} L^n .
\label{rho_n}
\end{equation}
The expectation values of all $\rho_n$ with $n=1,2, \dots, N-1$ form a complete
set of order parameters for all possible symmetry breaking patterns of $Z(N)$. 

In the large $N$ limit,  we use the notation $\theta_i = \theta (\frac{i}{N}-\frac12) =
\theta (x) $ with $-1/2 \leq x \leq 1/2$ \cite{Brezin:1977sv} and write
the Polyakov loop as
\begin{equation}
\rho_n = \frac{1}{N} \sum^N_{i=1} e^{in \theta_i}  \rightarrow  
\int^{\frac{1}{2}}_{-\frac{1}{2}} dx \exp \left[ i n \theta(x) \right] =  \int^\pi_{-\pi} d\theta \rho(\theta)  e^{i n \theta} ,
\label{rho_n_x}
\end{equation}
where we have introduced the eigenvalue density, $\rho(\theta)= dx/d\theta$, in the last expression.
The Polyakov loop thus becomes a functional of $\rho$.  
By definition the eigenvalue density has to satisfy two conditions,
the non-negativity 
 \begin{equation}
0 \leq \rho(\theta) , 
\label{nonnegative_condition}
\end{equation}
and the normalization 
\begin{equation}
1 = \int^{\pi}_{-\pi} d \theta \rho(\theta)\,. 
\label{normalization_condition}
\end{equation}
We will show in the next section that the non-negativity condition plays an essential
role for the GWW phase transition.

\subsection{Effective potential}

We construct the effective potential of the Polyakov loop near $T_d$ at large $N$ based on a version of the semi-classical method. The potential consists of two parts:
\begin{equation}
V_{\rm eff} (\rho_n)= V_{\rm pert} (\rho_n) + V_{\rm nonpert} (\rho_n), 
\end{equation}
where $V_{\rm pert}$ is the perturbative contribution, which breaks $Z(N)$ symmetry, and $V_{\rm nonpert}$ is the nonperturbative contribution, which keeps the theory in the confined phase below $T_d$.  The perturbative contribution is computed up to two
loops for any $N$ \cite{Dumitru:2013xna}: 
\begin{equation}
V_{\rm pert} = - \sum^{\infty}_{n=1}  d_n \left| \rho_n \right|^2 +  \mathcal{O}(\lambda^2) 
\;\;\;\;\; \mbox{with} \;\;\;\;\;
d_n = d_1 \frac{1}{n^d}
\label{V_pert}
\end{equation}
where $\lambda = g^2 N$ is the 't Hooft coupling, and $d$ is the number of  dimensions. 
The prefactor $d_1$ is positive and dimensionless, but the explicit form is irrelevant in this paper. 
The mass for each Polyakov loop $\rho_n$ is negative, so the perturbative effective potential breaks $Z(N)$ symmetry maximally. 
On the other hand, 
the leading-order contribution to the nonperturbative
part in the semi-classical approximation is unknown. It has to be such that it gives positive mass to every Polyakov loop below $T_d$ in order to prevent the spontaneous symmetry breaking.  
Therefore we consider the following form
\begin{equation}
V_{\rm nonpert} = \sum^\infty_{n=1} c_n \left| \rho_n \right|^2  + \dots\,, 
\label{V_np}
\end{equation}
where $c_n >d_n$ below $T_d$. 
For finite $N$, we only need the first $\left \lfloor{N/2}\right \rfloor $ terms, but for infinite $N$, it must contain infinite sum of the double-trace terms.   

We now make another assumption.  We assume that $Z(N)$ symmetry breaks completely at $T_d$ so that the phase transition is driven by the Polyakov loop $\rho_1$.  Lattice simulations in three and four dimensions at large $N$ support this assumption \cite{Lucini:2012gg}. 
We add a quartic coupling $b_1$ as a next-leading term, as well as the external field $h$ for $\rho_1$. 
We write the effective potential near $T_d$ as
\begin{equation}
V_{\rm eff} =  \sum^{\infty}_{n=1} a_n \left| \rho_n \right|^2 + b_1 \left( \left| \rho_1 \right|^2 \right)^2 -  h \left( \rho_1 +\rho^{*}_1 \right)  = \sum^{\infty}_{n=1} V_n (\rho_n).
\label{V_eff_h}
\end{equation}
where $a_n = c_n - d_n$.
This form is nothing but the sum of Landau free energy $V_n$ for each $\rho_n$,
where $V_{1} = a_1 \left| \rho_1 \right|^2 +b_1 \left( \left| \rho_1 \right|^2 \right)^2 - h \left(\rho_1 + \rho^*_1 \right) $ and
$V_{n>1} = a_n \left| \rho_n \right|^2$. 
Naively, it appears that each loop is independent and there is no interaction between $\rho_n$ 
and $\rho_m$ for $n\ne m$. The non-negativity condition (\ref{nonnegative_condition}), however,
constraints the eigenvalues of the Polyakov loop, and it couples all
$\rho_n$, as we show in the next two sections. 
 
Equation (\ref{V_eff_h}) reduces to the original model of Gross, Witten and Wadia when $b_1=0$, $a_n =c_n= 1/n$, and $h=1/g^2$.  The strong-coupling lattice gauge theory in mean-field approximation essentially reduces down to the model of Gross, Witten and Wadia, as well as the perturbation theory in $S^3 \times S^1$ \cite{Aharony:2003sx} although the perturbative contribution $-d_n$ is more complicated in this case.  In matrix models, both the Haar measure type ($c_n=1/n$) and the mass-deformation type ($c_n=1/n^2$) have been studied in \cite{Dumitru:2004gd} and \cite{Pisarski:2012bj}, respectively.

\section{Phase structure}

\begin{figure}[t]\centering
\begin{minipage}{.47\textwidth}
\subfloat[Without the external field, $h=0$.]{
\includegraphics[scale=0.4]{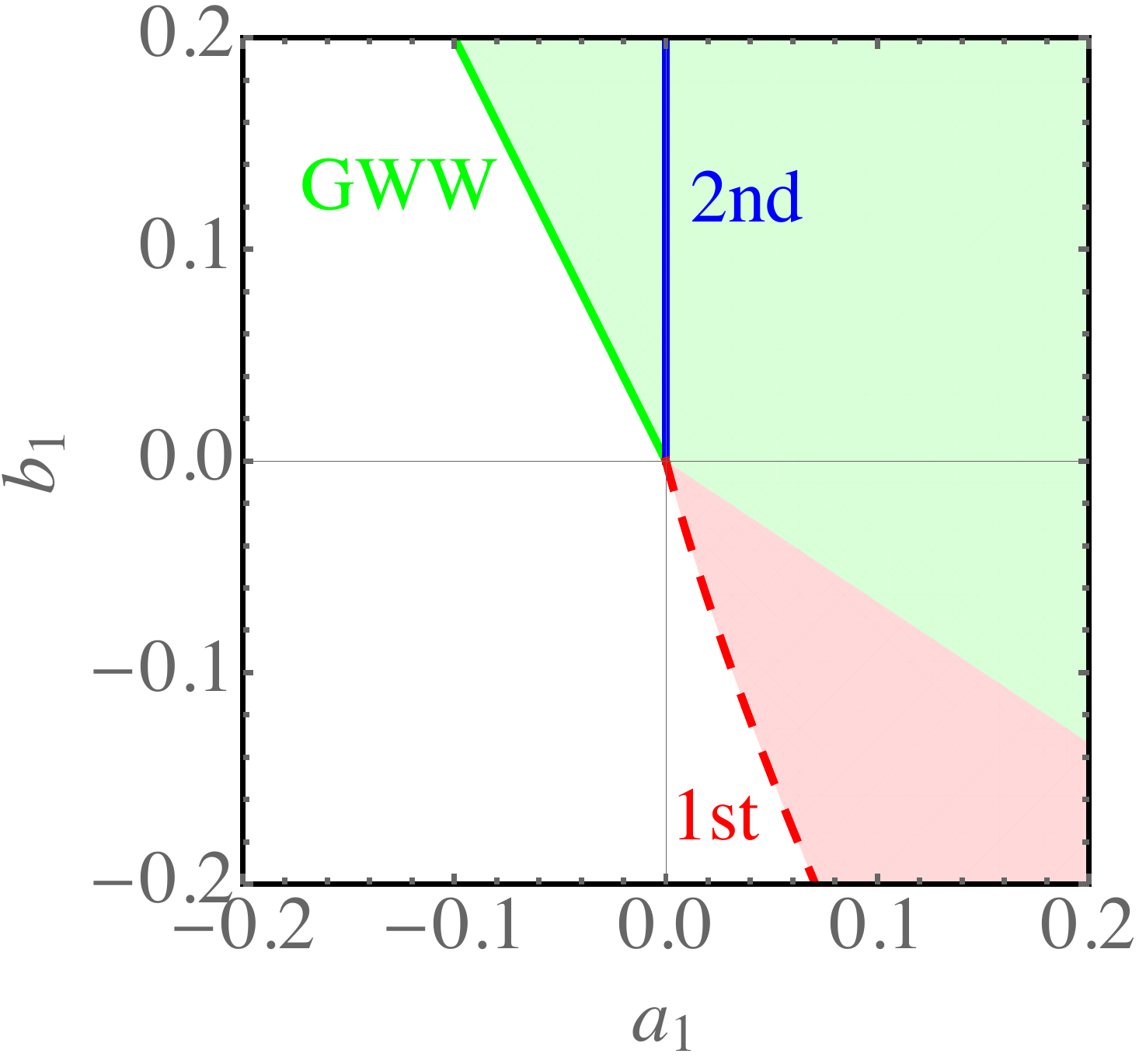}
\label{PhaseDiagram_2D}
}\end{minipage}
\begin{minipage}{.47\textwidth}
\subfloat[With the external field, $h \geq 0$.]{
\includegraphics[scale=0.5]{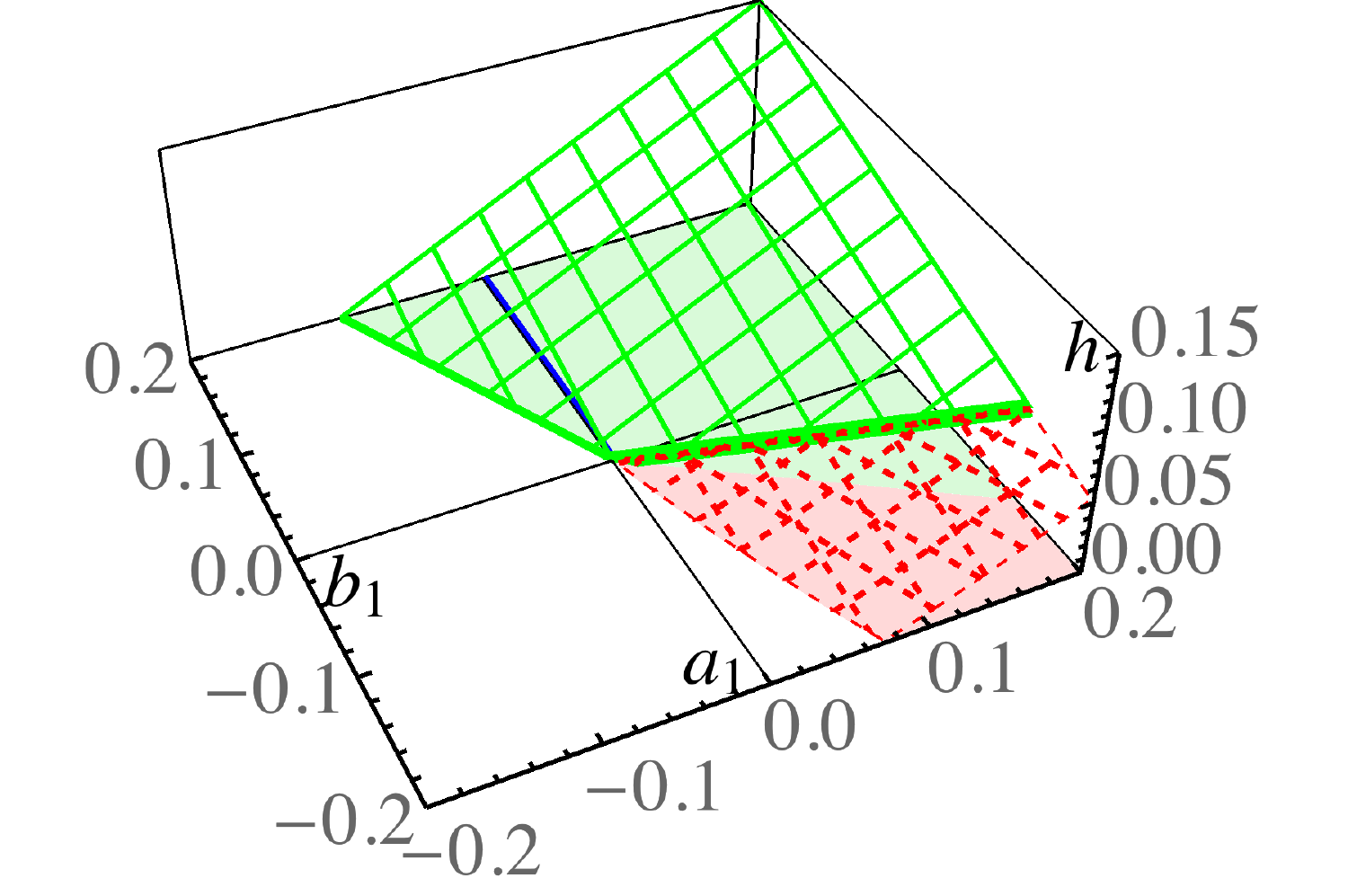}
\label{PhaseDiagram_3D}
}\end{minipage}
\caption{Phase diagrams of $SU(\infty)$ pure Yang-Mills constructed from
	Eq.~(\ref{V_eff_h}). The lines and meshed surfaces indicate phase
transitions: Solid lines are model independent while the dashed lines mildly
depend on the models. Here we have used the Vandermonde determinant type
($s=1$). The green solid and red dashed meshed surfaces are the surfaces of GWW and first-order phase transitions, respectively. The critical first order is located at the origin where all three phase transitions meet.
}
\label{PhaseDiagram}
\end{figure}

We construct a phase diagram based on Eq.~(\ref{V_eff_h}). We assume that the deconfining transition temperature $T_d$ defined at $h=0$ is driven by the first Polyakov loop $\rho_1$, so that $Z(N)$ symmetry is completely broken at $T_d$. We further assume that 
$b_1$ is small and constant near $T_d$ and also that
$a_{n>1}>0$ near $T_d$ so that the higher corrections for $\rho_{n>1}$ are not included.  We otherwise keep the coefficients $a_n$ arbitrary in this section. The three parameters in $V_1(\rho_1)$, i.e., $a_1$, $b_1$ and $h$, determine the phase structure as shown in Fig.~\ref{PhaseDiagram} 

In the confined phase, all $a_n$ are positive so that there is no spontaneous symmetry breaking of $Z(N)$. The confined phase is the region to the right of the 1st and 2nd order phase transition lines in Fig.~\ref{PhaseDiagram_2D}. As we increase the external field $h$, $Z(N)$ symmetry is explicitly broken, and the Polyakov loop $\rho_1$ grows monotonically with $h$, while all other Polyakov loops $\rho_{n>1}$ stay zero. The eigenvalue density can be then written as a Fourier series in terms of the moments
\begin{equation}
\rho(\theta) = \frac{1}{2\pi} \left( 1 + 2 \rho_1(h) \cos \theta \right)
\label{rho_belowGWW}
\end{equation}
where $-\pi \leq \theta \leq  -\pi$. 
There is a critical value of the external field $h_c$ where $\rho_1 (h_c) = 1/2$, above which this solution violates the nonnegative condition (\ref{nonnegative_condition}). 
Writing $\delta h = h- h_c$, we can express the free energy as \cite{Nishimura:2017crr}
\begin{equation}
\displaystyle
F(h) = f_{\rm reg}(h) +
\left\{ 
\begin{array}{lcl}
0 & \mbox{for} & \delta h \leq 0
\\ 
\displaystyle
v \delta h^r + \mathcal{O}(\delta h^{r+1})  & \mbox{for} & \delta h > 0  \; ,
\end{array}
\right.
\label{F}
\end{equation}
where $v$ is an irrelevant constant, 
and $f_{\rm reg}$ is a smooth function of $h$. We argue in \cite{Nishimura:2017crr} that the exponent $r$ is larger than two for any coefficients $a_n$, and thus the singular point corresponds to a continuous phase transition whose order is larger than second. 
We call this transition a generalized Gross-Witten-Wadia (GWW) transition. 
The original model of Gross, Witten, and Wadia \cite{Gross:1980he, Wadia:1980cp}, 
gives rise to the Vandermonde-determinant-type potential where $a_n = 1/n$, and
the exponent  $r$ is three, i.e. the transition is third order.

The surface of the GWW phase transition where $\rho_1$ becomes $1/2$ can be found by expanding $V_1$ around $\rho_1 = 1/2$.  We plot the GWW surface in Fig.~\ref{PhaseDiagram_3D} with the green solid meshed surface.  The GWW surface is independent of the detail of the coefficients for the double trace terms.  When $b_1$ is negative and $\left| b_1 \right|$ sufficiently large, the phase transition becomes first order.  The location of the first-order phase transition depends on the details of the coefficients $a_n$: here we plot the surface of first-order phase transition using the model based on the Vandermonde determinant ($s=1$) indicated by the red dashed meshed surface in Fig.~\ref{PhaseDiagram_3D}. 

Fig.~\ref{PhaseDiagram_2D} is the phase diagram with zero external field.  The green and red shaded regions are the projections of the GWW and first-order phase transition surfaces onto the $h=0$ plane, respectively.  There is a second-order phase transition line where $a_1=0$ and $b_1>0$. At the origin $a_1= b_1 = h = 0 $, the first, second and higher order phase transition lines meet.  At this point, the Polyakov loop $\rho_1$ jumps from $0$ to $1/2$, as is typical of a first order phase transition, while the mass associated with $\rho_1$ becomes zero, as is typical of a second order phase transition. This point is termed as ``critical first order" in \cite{Dumitru:2003hp}.

\section{Models}
\label{Models}

We now model the coefficients of effective potential $a_n$ in Eq.~(\ref{V_eff_h}) and confirm the general argument in the previous section.
The coefficients of the double trace terms, $a_n=c_n -d_n$,
consist of two parts: the perturbative contribution $-d_n$ (\ref{V_pert})
and the unknown nonperturbative contribution $c_n$. In the confined phase, $c_n $
has to be larger than $d_n$ for all $n$.  A
simple choice is to use
\begin{equation}
c_n  = c_1(T/T_d)  \frac{1}{ n^{s}},
\label{c_n}
\end{equation}
where $c_1$ is a dimensionless function of $T/T_d$. We set $s \leq d$ and $c_1>
d_1$ below $T_d$, so that the phase transition happens at $c_1 = d_1$ for nonnegative $b_1$. 
In this paper, we take $s=1,2,3,$ and $4$.  $s=1$ and $2$ are the type of
confining potentials used in phenomenological models of QCD at finite
temperature based on the Haar measure and the mass deformation, respectively
\cite{Meisinger:2001cq}. 
On the other hand, the confining potential with $s=4$ is based
on the the ghost dominance in the infrared regime
\cite{Zwanziger:2001kw}, which gives rise to the so called
inverted GPY-Weiss potential. Alternatively, one can think of the model as the mass deformation potential in $2s$-dimensions or the inverse GPY-Weiss potential in $s$ dimensions, but we do not specify dimensions and consider $s=1,2,3$, and $4$ with $d \geq s$ as simple examples.

\begin{figure}[t]\centering

\begin{minipage}{.47\textwidth}
	\subfloat[$\rho=dx/d\theta$ as a function of $\theta$]{
\includegraphics[width=0.97\textwidth]{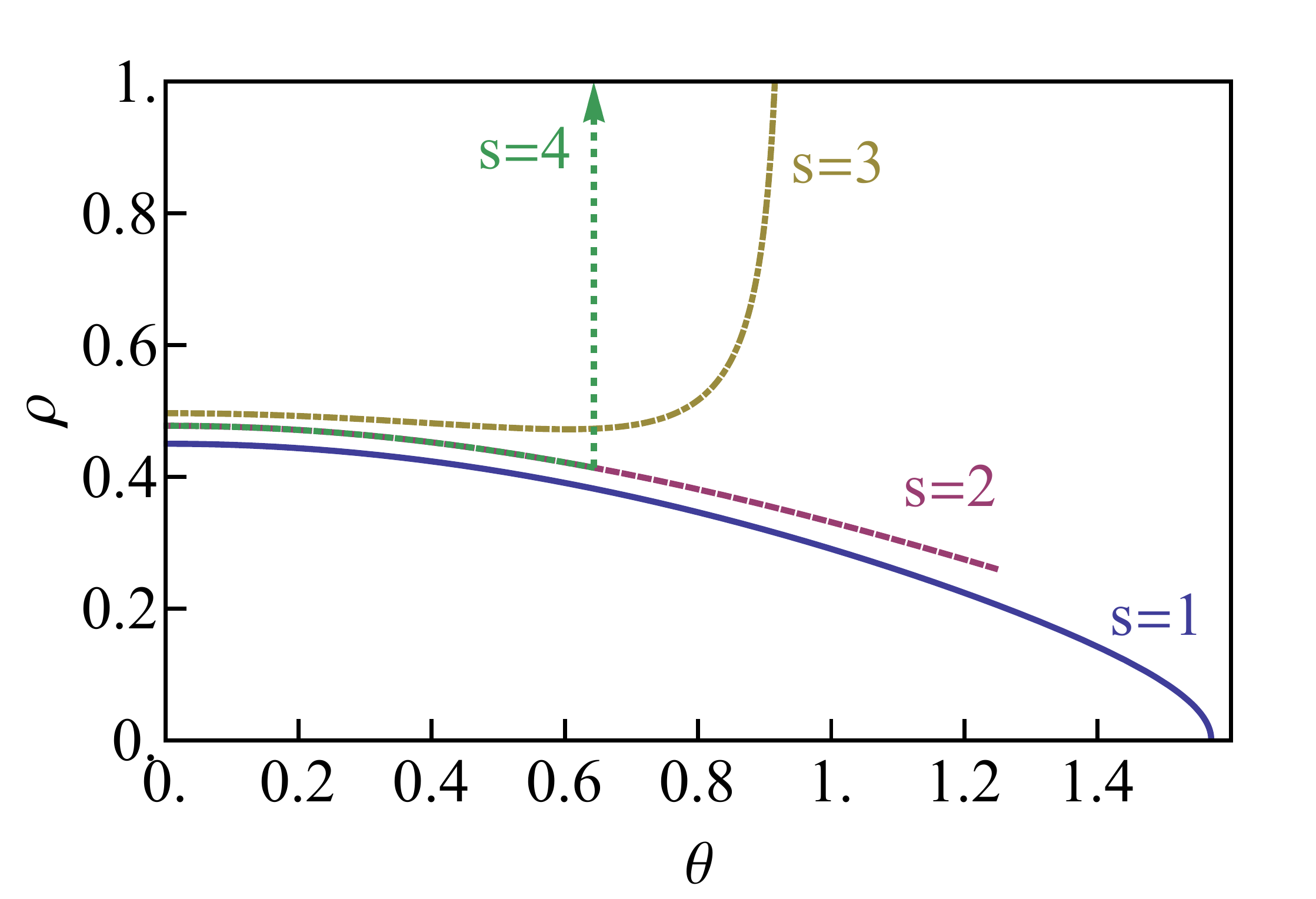}
\label{Distribution_AboveGWW}
}\end{minipage}
\; \;
\begin{minipage}{.47\textwidth}
\subfloat[$x$ as a function of $\theta$]{
\includegraphics[width=\textwidth]{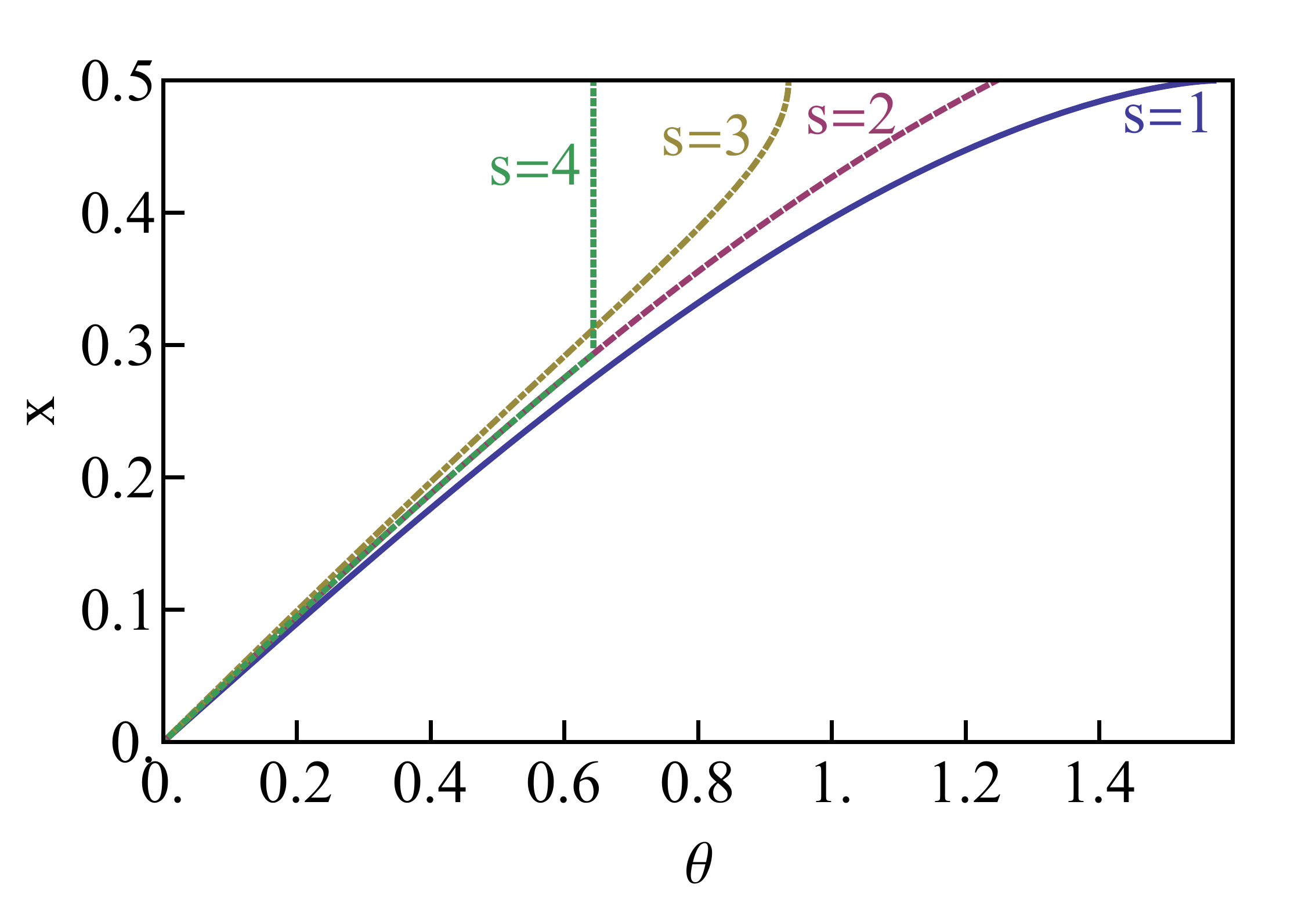}
\label{theta_vs_x_aboveGWW}
}\end{minipage}
\caption{The eigenvalue density above the GWW point, $h=1$. The eigenvalue density changes depending on the coefficients of the double trace terms, $s=1,2,3,$ and $4$ as shown in the left plot. 
The right plot shows $x$ as a function of $\theta$. The eigenvalues for $s=4$ pile up at the end point for $\theta(x)$ with $x \geq x_0$.  As a result, the eigenvalue density becomes the delta function at the endpoint $\pm \theta(x_0)$ as depicted by the arrow in the left plot. Note that $\rho(-\theta) = \rho(\theta)$ and $x(-\theta) =- x (\theta)$.
}
\label{Distribution}
\end{figure}

In this paper, we set $d_1=0$ and $b_1=0$ in Eq.~(\ref{V_eff_h}) and solve the model exactly.   
The exact solution can be then used to construct the full potential including $b_1$ and $d_1$ using the Legendre transform \cite{Nishimura:2017crr} following \cite{Dumitru:2004gd}.  
The equation of motion is then
\begin{eqnarray}
h \sin \theta
&=&
\int^{\pi}_{-\pi} d\theta' \rho(\theta')  \sum^{\infty}_{n=1}  \frac{\sin(n \theta - n \theta')}{n^{s-1}} \, ,  
\label{EoM_GWW_Integral}
\end{eqnarray}
where we have set $c_1=1$ without loss of generality. We solve this equation under the two constraints, Eqs.~(\ref{nonnegative_condition}) and
(\ref{normalization_condition}).  
Below the GWW point $h \leq h_c = 1/2$, the solution for the equation of motion is given in Eq.~(\ref{rho_belowGWW}) with $\rho_1=h$ and $\rho_{n>1} =0$. As discussed before, the solution is independent of the model. The free energy is given as $F= c_1 \rho^2_1(h) - 2 h \rho_1(h) =-\frac{1}{4} - \delta h - \delta h^2 $ where $\delta h = h -h_c$. 

The eigenvalue density above the GWW point depends on models, but it can be solved exactly for $s=1,2,3$ and $4$. Above the GWW point, the eigenvalue density develops a gap at the endpoints for all cases. It is given as
\begin{eqnarray}
s=1,3: \,\,\,\,\, \rho (\theta) &=&
C_1 \cos \frac{\theta}{2} \left( \sin^2 \frac{\theta_0}{2} - \sin^2 \frac{\theta}{2} \right)^{1/2} + C_2 \cos^3 \frac{\theta}{2} \left( \sin^2 \frac{\theta_0}{2} - \sin^2 \frac{\theta}{2} \right)^{-1/2} 
\\
s=2,4: \,\,\,\,\, \rho (\theta) &=&
\left(1-x_0\right) \frac{\delta(\theta-\theta_0)+\delta(\theta+\theta_0)}{2}+
\frac{1}{2 \pi }\left(1 + 2 h \cos \theta \right)
\label{rho_odd}
\end{eqnarray}
where $C_1 + C_2 = 2 h / \pi$.
The density is defined only in the interval of $-\theta_0 \leq \theta \leq \theta_0$, and it is zero otherwise.
The endpoint $\theta_0$ is given implicitly as a function of $h$:
\begin{equation}
h = \frac{1}{2 \sin^2 \frac{\theta_0}{2}}, \;
\frac{\pi - \theta_0 }{2 \sin (\pi - \theta_0)} , \;
-\frac{\ln (\sin \frac{\theta_0}{2})}{1-\sin^2 \frac{\theta_0}{2}}, \;
\frac{\left(\pi - \theta_0 \right)^3  }{6 \sin \theta_0+6 \left( \pi -\theta_0 \right) \cos \theta_0}
\end{equation}
for $s=1,2,3$ and $4$, respectively. The solution is the minimum of the potential if $C_1 = 2h / \pi$ for $s=1$, while $x_0=1$ for $s=2$. On the other hand, the solution satisfies the equation of motion only if $ C_1 = \left( -1 + 3 h + h  \cos \theta_0 \right) / \left(\pi+\pi \cos \theta_0 \right)$ for $s=3$, while  $x_0 = \left( \theta_0 + 2 h \sin \theta_0 \right) /\pi $
for $s=4$.  
The eigenvalue density above the GWW point is plotted in Fig.~\ref{Distribution_AboveGWW} when $h=1$.

For $s=4$, the eigenvalues $\theta(x)$ with $x_0/2 \leq x  \leq 1/2$ become a single value $\theta_0=\theta(x_0/2)$ as shown in Fig.~\ref{theta_vs_x_aboveGWW}. Therefore $\theta$ is no longer an injective function of $x$, and the density becomes a delta function at the endpoints, as depicted by the arrow in Fig.~\ref{Distribution_AboveGWW}. 
Physically, the eigenvalue repulsion weakens as $s$ increases, and the value of $\theta_0$ for a fixed value of $h$ becomes smaller. When $s = 4$ the repulsion is so weak that the eigenvalues pile up at the endpoints, and the density becomes delta function.

The free energy can be computed using the solutions above. It is given as in Eq.~(\ref{F}) with $f_{\rm reg} = -\frac{1}{4}-\delta h - \delta h^2$ and $v = 4/3 \; , \; 128 \sqrt{3}/(35 \pi) \; , \; 8/3 \; , \; 2560 \sqrt{5}/ ( 567 \pi) $ for $s=1,2,3$ and $4$, respectively. The exponent is given as
\begin{equation}
r = \frac{5+s}{2} .
\end{equation}
The order of the discontinuity of the free energy with respect to $h$
depends upon $s$: for $s=1$ the third
derivative is discontinuous; for $s=2$ and $3$, the fourth derivative; and
for $s=4$, the fifth derivative.


\section{Conclusions}

We have considered an effective potential of the Polyakov loop in $SU(N)$
Yang-Mills theory at large $N$ using a version of semi-classical argument.
According to perturbation theory, only the double trace of the Polyakov loop is
present up to two loops.  We assume that the double-trace terms dominate near
the phase transition and argue that the effective potential takes the form
given in Eq.~(\ref{V_eff_h}) in the presence of the background field $h$
coupled to the Polyakov loop $\rho_1$.  We have shown that there are at least
three different types of phase transitions as shown in Fig.~\ref{PhaseDiagram}.
Only the location of the first-order phase transition depends on the explicit form of the 
coefficients, while the locations of the other phase transitions are model independent.  There is a large region in the parameter space where the phase transition
is continuous but the order of phase transition is third or larger as depicted in the green solid meshed surface in Fig.~\ref{PhaseDiagram_3D}. We have called the surface of the phase transition the Gross-Witten-Wadia (GWW) surface; anywhere on this surface the Polyakov loop $\rho_1$ is $1/2$.  Below the GWW point, the effective potential is simply a sum of simple Landau free energy for each $\rho_n$. At the GWW point, the Polyakov loop $\rho_1$ becomes $1/2$ and the simple Landau theory breaks down due to the nonnegativity constraint for the eigenvalue density.   We have confirmed this general argument using specific models in Sec.~\ref{Models}. Both the free energy and the eigenvalue density drastically change above the GWW point. Observing these behavior in lattice simulations at large $N$ would be a benchmark for the GWW phase transition.

\end{document}